\documentclass[a4paper,fleqn,usenatbib]{mnras}

\usepackage{txfonts}

\usepackage[T1]{fontenc}
\usepackage{ae,aecompl}

\usepackage[toc,page]{appendix}
\usepackage{natbib}
\usepackage{amssymb}
\usepackage{epsf}
\usepackage{pstricks}
\usepackage{color}
\usepackage{graphicx}
\usepackage{slashed}
\usepackage{relsize}
\usepackage{morefloats}
\usepackage{multirow}
\usepackage{array}
\usepackage{hyperref}
\usepackage[justification=centering]{caption}
\usepackage{url}

\title[Prospects for Discovering Pulsars in Future Continuum Surveys Using Variance Imaging]{Prospects for Discovering Pulsars in Future Continuum Surveys Using Variance Imaging}

\author[S. Dai et al.]{
S. Dai$^{1}$\thanks{E-mail: shi.dai@csiro.au},
S. Johnston$^{1}$,
G. Hobbs$^{1}$\\
$^{1}$CSIRO Astronomy and Space Science, Australia Telescope National Facility, Box 76 Epping NSW 1710, Australia\\
}

\date{Accepted XXX. Received YYY; in original form ZZZ}

\pubyear{2017}

\begin{document}
\label{firstpage}
\pagerange{\pageref{firstpage}--\pageref{lastpage}}
\maketitle

\begin{abstract}
In \citet{djb+16} we developed a formalism for computing variance
images from standard, interferometric radio images containing time
and frequency information. Variance imaging with future radio continuum
surveys allows us to identify radio pulsars and serves as a complement to
conventional pulsar searches which are most sensitive to strictly periodic signals.
Here, we carry out simulations to predict the number of pulsars that we can 
uncover with variance imaging on future continuum surveys. We show that the 
Australian SKA Pathfinder (ASKAP) Evolutionary Map of the Universe (EMU) survey 
can find $\sim30$ normal pulsars and $\sim40$ millisecond pulsars (MSPs) over 
and above the number known today, and similarly an all-sky continuum survey with 
SKA-MID can discover $\sim140$ normal pulsars and $\sim110$ MSPs with this technique.
Variance imaging with EMU and SKA-MID will detect pulsars with large duty cycles 
and is therefore a potential tool for finding MSPs and pulsars in relativistic 
binary systems.
Compared with current pulsar surveys at high Galactic latitudes in the southern 
hemisphere, variance imaging with EMU and SKA-MID will be more sensitive, and 
will enable detection of pulsars with dispersion measures between $\sim10$ and 100\,cm$^{-3}$\,pc.
\end{abstract}




\section{Introduction}

The next-generation of radio continuum surveys will provide us with
ultra-deep images of large swathes of the sky~\citep[e.g.,][]{nha+11,ps15}. 
As a result, millions of radio sources will be detected, a tiny (but significant)
fraction of which will be radio pulsars. How can we distinguish radio pulsars from all the other sources?
In \citet{djb+16} (hereafter D16), we proposed to use the variance of
flux densities caused 
by diffractive interstellar scintillation to distinguish pulsars from other 
radio sources. In conjunction with the conventional all-sky blind searches for
pulsars, this technique has the potential to uncover pulsars missed
by the standard techniques.
This is because variance imaging can detect pulsars independent of
their pulse width. This is not the case with conventional pulsar searches
which are most sensitive to narrow pulse-widths. A variety of effects
can broaden the pulse profile, including dispersion-measure
(DM) smearing, scattering and orbital modulation of spin periods.
This makes it possible to identify sub-millisecond pulsars, pulsar-blackhole 
systems and pulsars in the Galactic centre in continuum surveys.

However, variance imaging also has its limitations.
In D16, we investigated the properties of variance imaging and showed that 
the sensitivity peaks for pulsars whose scintillation bandwidth and 
time-scales are close to the channel bandwidth and subintegration time of
a particular continuum image. Therefore, for a given continuum survey with
certain total bandwidth and total integration time it is best
to retain a high number of frequency channels across the band, and a high number of sub-integrations.
However, as we increase the time and frequency resolution, the overall sensitivity of variance imaging will decrease 
because the noise level in each channel and subintegration increases.
This leads to a `sweet-spot' in channels and subintegrations. In turn
this limits the volume in which pulsars can be found, because more distant
pulsars have narrower scintillation bandwidths.
We also show in D16 that the sensitivity of variance 
images will be lower than that of Stokes I images.

In this paper, we carry out pulsar population simulations and predict the 
number of pulsars that can be detected with variance imaging for
future continuum surveys.  We also investigate distributions and properties of 
pulsars detected in variance images and compare them with those of pulsars 
found through conventional searches.  In Section~\ref{psr_sim}, we describe 
the setup of pulsar population simulations, including pulsar distributions and 
Galactic electron-density models.
In Section~\ref{periodic_search}, we discuss the sensitivity of periodic 
search and variance imaging.
In Section~\ref{var_image}, we predict the number of pulsars that can be 
found with variance imaging for future continuum surveys. 
We discuss our results and conclude in Sections~\ref{discussion} and \ref{conclusion}.


\section{Pulsar population simulations}
\label{psr_sim}

We define millisecond pulsars (MSPs) as pulsars with spin period $P<20$\,ms
and `normal' pulsars with periods larger than this.
According to previous studies~\citep[e.g.,][]{lbb+13,lem+15}, 
MSPs and normal pulsars have different spin period distributions and Galactic scale heights,
and therefore we simulate the MSP and the normal pulsar populations separately.

To simulate pulsar populations, we used the PSRPOPPY software package~\citep{blr+14}  
which is based on the PSRPOP package~\citep{lfl+06}. The {\sc populate} module was 
used to simulate pulsars and pulsar parameters were drawn from specified distributions.
The {\sc dosurvey} module of PSRPOPPY was used to simulate existing pulsar surveys.

\subsection{Normal pulsars}

For the radial distribution of normal pulsars, we used the model of \citet{yk04}, 
\begin{equation}
	\rho(r)\propto\left(\frac{r+R_{1}}{R_{\odot}+R_{1}}\right)^{a}\exp{\left[-b\frac{r-R_{\odot}}{R_{\odot}+R_{1}}\right]},
\end{equation}
where $\rho(r)$ is the surface density at galactocentric radius $r$; $R_{\odot}$ is the 
Sun to the Galactic center distance; $R_{1}=0.55$\,kpc, $a=1.64$ and $b=4.01$ are model parameters.
The distribution of normal pulsars in Galactic $z$ coordinates was approximated by a two-sided 
exponential~\citep{lfl+06},
\begin{equation}
	N(z)\propto\exp{\left(\frac{-|z|}{E}\right)},
\end{equation}
where $E=0.33$\,kpc is the scale height.
We used a log-normal distribution of pulse period~\citep{lfl+06},
\begin{equation}
	f(P)\propto\exp{\left[-\frac{(\log{P}-I)^2}{2J^2}\right]},
\end{equation}
where $I=2.7$ is the mean and $J=-0.34$ is the standard deviation.
The beaming model described in \citet{slk+09} was used. We assumed a mean spectral 
index of $-1.6$ and a standard deviation of 0.35~\citep{lyl+95}.
Pulsar luminosities were drawn from a log-normal distribution with a mean of 
$\langle \log_{10}L\rangle = -1.1$ and a standard deviation of $\sigma_{\log_{10}L}=0.9$~\citet{fk06}.

We normalised the total number of potentially observable pulsars in the 
Galaxy to give a yield of 1038 pulsars for the Parkes Multibeam (PM) 
Pulsar Surveys~\citep{mlc+01}. This results in an underlying sample 
of 120\,000 Galactic normal pulsars, which is consistent with predictions from \citet{fk06}. 

\subsection{Millisecond pulsars}

We assumed the same spectral index and radial distribution 
for MSPs as those of normal pulsars. The distribution of MSPs in Galactic $z$ coordinates 
was approximated by a two-sided exponential with a scale height of 0.5\,kpc~\citep{lbb+13}.
For the pulse period distribution of MSPs, we used a log-normal distribution following \citet{lem+15}, 
\begin{equation}
	f(P)\propto\frac{1}{P}\exp{\left[-\frac{(\ln{P}-E)^2}{2F^2}\right]}
\end{equation}
where $E=1.5$ and $F=0.58$.
We used the luminosity distribution of MSPs obtained by \citet{lbb+13}, 
\begin{equation}
	N\propto\left(\frac{L}{\rm{mJy}\ \rm{kpc^2}}\right)^{-1.45},
\end{equation}
with luminosity $L$ cuts at 0.1\,mJy\,kpc$^2$ and 100\,mJy\,kpc$^2$. As discussed in \citet{lbb+13}, 
this distribution is significantly steeper than that of normal pulsars and does not represent
the luminosity distribution of MSPs below the luminosity value for the weakest pulsar in 
their sample. For the purpose of this paper, considering the limited sensitivity of 
variance imaging, our results will not be affected by the population of extremely weak MSPs.

We normalised the total number of potentially observable MSPs in the Galaxy to give 
48 MSPs for the intermediate latitude part of the southern High Time Resolution Universe (HTRU) 
survey~\citep{kjv+10}. This results in an underlying sample 
of 41\,000 Galactic MSPs. 

\subsection{Galactic electron-density model}

\citet{ymw17} model was used to estimate DMs for given pulsar Galactic longitudes and latitudes and their distances. 
The \citet{ymw17} model also gives us pulsar scattering time-scales ($\tau_{\rm{sc}}$)
using the relation obtained by \citet{kmn+15} for the DM dependence of observed $\tau_{\rm{sc}}$
values (in units of seconds and scaled to 1\,GHz assuming $\tau_{\rm{sc}}\approx\nu^{-4}$, 
$\nu$ is the frequency in units of GHz):
\begin{equation}
	\tau_{\rm{sc}}=4.1\times10^{-11}\ \rm{DM}^{2.2}(1.0+0.00194\ \rm{DM}^{2}).
\label{tau_sc}
\end{equation}
We estimate the scintillation bandwidth using
\begin{equation}
	2\pi\delta\nu_{\rm{DISS}}\tau_{\rm{sc}} = C_1,
\end{equation}
where $C_1=1.16$ for a uniform medium with a Kolmogorov wavenumber spectrum~\citep[e.g.,][]{cr98}.
The scintillation time-scale is estimated using Eq. 13 of \citet{cr98}  
\begin{equation}
	\tau_{\rm{DISS}}=A_{\rm{ISS}}\frac{\sqrt{D\delta\nu_{\rm{DISS}}}}{\nu V_{\rm{ISS}}},
\end{equation}
where $A_{ISS}=2.53\times10^4$\,km\,s$^{-1}$, $D$ is the pulsar distance and $V_{\rm{ISS}}$
is the speed of interstellar diffraction pattern relative to the Earth in units of km\,s$^{-1}$.
We rescale the scintillation bandwidth and time-scale to other observing frequencies
using $\tau_{\rm{DISS}}\propto\nu^{6/5}$ and $\delta\nu_{\rm{DISS}}\propto\nu^{22/5}$~\citep[e.g.,][]{cr98}.
Since velocities of the observer and the ISM are normally negligible compared with the 
transverse velocity of pulsars $V_{\rm{p}\perp}$, we simply rescale the scintillation 
time-scale to different pulsar transverse velocities using $\tau_{\rm{DISS}}\propto100/V_{\rm{p}\perp}$.
For pulsar transverse velocities, we used an exponential distribution with a mean velocity 
of $\langle V_{\rm{p}\perp}\rangle=180$\,km\,s$^{-1}$~\citep{fk06}.

\section{Sensitivity of pulsar searches}
\label{periodic_search}

\subsection{Periodic search}

Conventional pulsar surveys search for strictly periodic signals.
Generally, pulsars have intrinsically narrow pulses, but 
smearing caused by interstellar dispersion, scattering, and modulation of 
orbital period can all significantly increase
the pulse width. The detection threshold of the apparent flux density 
for pulsar surveys can be calculated as
\begin{equation}
	S_{\rm{min}}=\frac{S/N_{\rm{min}}(T_{\rm{rec}}+T_{\rm{sky}}+T_{\rm{CMB}})}{G\eta\sqrt{n_{\rm{pol}}t_{\rm{int}}\Delta\nu}}\sqrt{\frac{\delta}{1-\delta}}\quad\rm{mJy},
\label{sensitivity_periodic}
\end{equation}
where $S/N_{\rm{min}}$ is the threshold signal-to-noise ratio ($S/N$), $T_{\rm{rec}}$, 
$T_{\rm{sky}}$ and $T_{\rm{CMB}}$ are the receiver, sky and cosmic microwave background (CMB) 
temperature (K), $G$ is the telescope antenna gain (K/Jy), $\eta$ is a survey-dependent 
constant ($\le1$) which accounts for losses in sensitivity due to, e.g. sampling and 
digitisation noise, $n_{\rm{pol}}$ is the number of polarisation, $t_{\rm{int}}$ is the 
integration time (second) and $\Delta\nu$ is the observing bandwidth (MHz). The 
pulse duty cycle is defined as
\begin{equation}
	\delta=W_{\rm{eff}}/P,
\end{equation}
where $W_{\rm{eff}}$ is the effective pulse width and $P$ is the pulse period.
The effective pulse width is given by
\begin{equation}
W_{\rm{eff}}=\sqrt{W_{\rm{int}}^2+t_{\rm{samp}}^2+\Delta t^2+\tau_{\rm{sc}}^2},
\end{equation}
where $W_{\rm{int}}$ is the intrinsic pulse width, $t_{\rm{samp}}$ is the sampling 
time. $\Delta t$ is the dispersive 
smearing time across a frequency channel of bandwidth $\Delta\nu_{\rm{chn}}$ (MHz)
at frequency $\nu$ (MHz) estimated as 
\begin{equation}
	\Delta t=8.3\times10^{6}\ \rm{ms}\times\rm{DM}\times\frac{\Delta\nu_{\rm{chn}}}{\nu^{3}}.
\end{equation}
The pulse smearing due to the scattering by the ionised interstellar medium, $\tau_{\rm{sc}}$,
can be estimated using Eq.~\ref{tau_sc}.

\begin{figure}
\center
\includegraphics[width=3.5in]{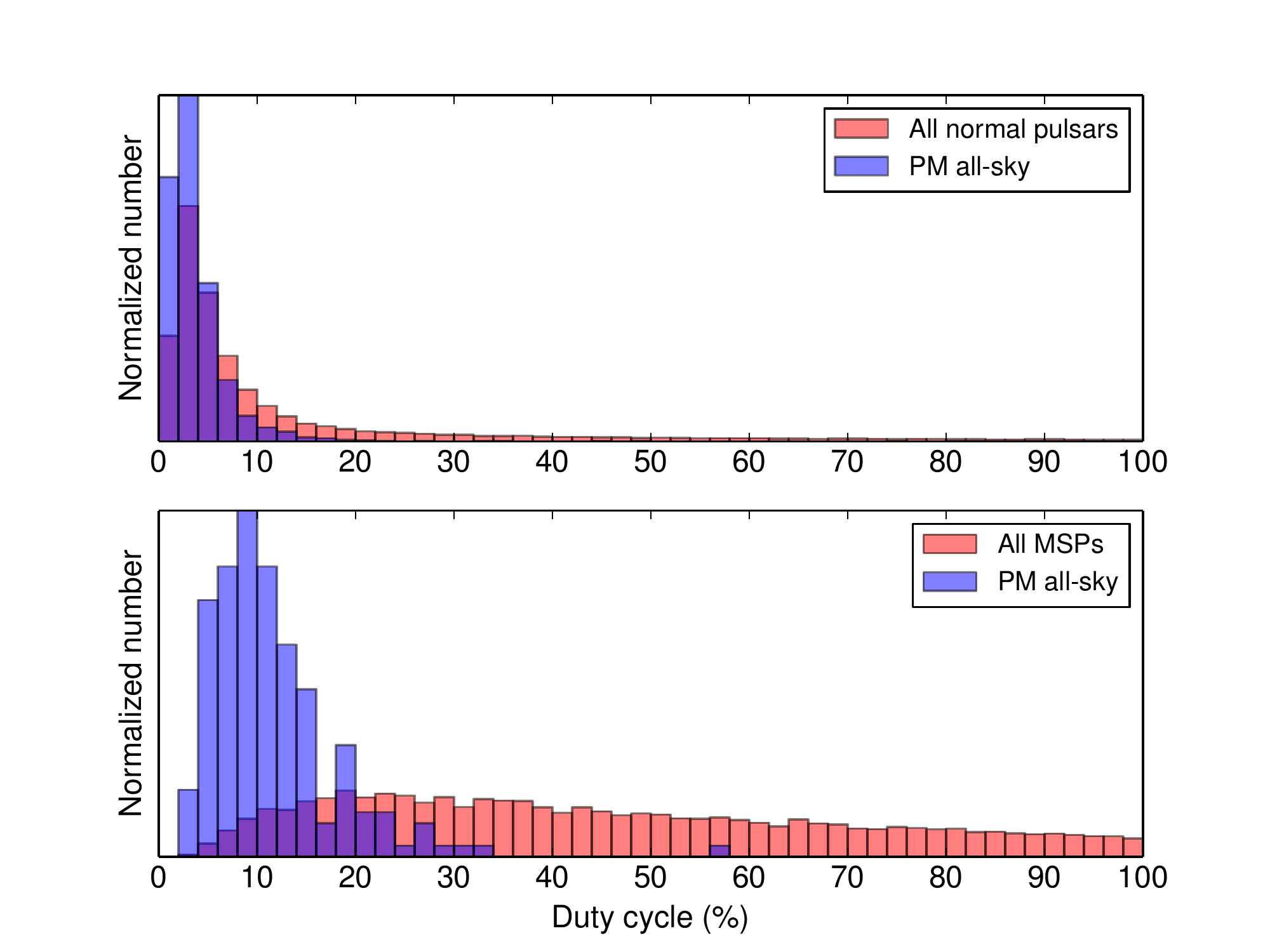}
\caption{Histograms of duty cycles for simulated pulsars. Top panel: 120\,000 Galactic normal 
pulsars (red) and 1980 pulsars detected with the simulated PM all-sky survey. Bottom panel: 41\,000 MSPs 
(red) and 175 MSPs detected with the simulated PM all-sky surveys.}
\label{duty}
\end{figure}

From Eq.~\ref{sensitivity_periodic} we can clearly see that as the duty cycle increases the
sensitivity of periodic search decreases, and therefore periodic searches will be biased 
towards low duty cycle pulsars and tend to miss pulsars smeared by dispersion, scattering 
and modulation of orbital period.
In Fig.~\ref{duty}, we demonstrate this by simulating 
surveys and pulsar populations using distributions described in Section~\ref{psr_sim}. 
The pulsar survey we simulated has identical parameters as the PM Pulsar Survey but with a 
larger sky coverage (declination $<+30^{\circ}$) and we call it PM all-sky survey. 
Fig.~\ref{duty} shows that periodic searches are missing large duty-cycle pulsars and 
this is particularly significant for MSPs since they have much shorter spin periods 
than normal pulsars and are therefore heavily affected by dispersive and scattering smearing.

We note that in these simulations we did not take the smearing caused by the modulation of 
orbital periods of MSPs into account. For MSPs in compact binary systems, the periods 
will be significantly modulated if the observing time is longer than the orbital period, 
which will cause smearing of the pulse profile and increase the duty cycle if such 
modulations are not taken care of. Therefore, the sensitivity of conventional periodic 
searches dramatically decreases for MSPs in compact binary systems. Although various 
methods have been proposed to minimise the effects of orbital modulations~\citep[e.g.,][]{rce03}, 
it is still challenging to search for MSPs in binary systems.

\subsection{Variance imaging}
\label{var_sensitivity}

Statistics and simulations of detecting pulsars in variance images have been discussed 
in detail in D16. For a frequency resolution of $\delta\nu$ and a time 
resolution of $\delta t$, variance imaging will be most sensitive to pulsars with 
scintillation bandwidths of $\delta\nu_{\rm{DISS}}\approx\delta\nu/2$ and scintillation 
time-scales of $\delta t_{\rm{DISS}}\approx\delta t/2$. 
To predict the number of pulsars that can be discovered by future surveys 
with variance imaging, we calculate sensitivities as a function of scintillation 
bandwidths and time-scales for a given survey with a total bandwidth of $B$ and an 
integration time of $T$. For each scintillation bandwidth $\delta\nu_{\rm{DISS}}$ and 
time-scale $\delta t_{\rm{DISS}}$, we use a channel number of $N_{\rm{f}}$ and a subintegration 
number of $N_{\rm{t}}$ so that $\delta\nu=B/N_{\rm{f}}=2\times\delta\nu_{\rm{DISS}}$ and 
$\delta t=T/N_{\rm{t}}=2\times\delta t_{\rm{DISS}}$. This corresponds to the ``matched filter'' 
for different scintillation bandwidths and time-scales and gives us the highest sensitivity.

Under such assumptions, the number of samples in the dynamic spectrum matches 
the number of independent scintles. Each sample in the 
dynamic spectrum can then be simulated as the sum of two Gaussian random 
variables (for details see D16).
We first determine the detection threshold in the images such that
it yields a five per cent false alarm probability.
Then we determine the sensitivity for which 
80 per cent of the measurements exceed the detection threshold.
This sensitivity is a function of the scintillation bandwidth and time-scale (see Fig.~5 in D16).

\section{Detecting pulsars in variance images}
\label{var_image}

We simulate populations of normal pulsars and MSPs in the Galaxy and estimate their 
scintillation bandwidth and time-scale as described in Section~\ref{psr_sim}. 
For a given continuum survey, the sensitivity of variance imaging as a function of  
scintillation bandwidth and time-scale can be estimated following Section~\ref{var_sensitivity}. 
If the pulsar flux density is above the sensitivity corresponding to its 
scintillation bandwidth and time-scale, we call it as a detection, which has a five per 
cent false alarm probability and 80 per cent detection probability. 
In this way we identify all pulsars that can be detected by variance imaging with 
a given continuum survey.

Many of these pulsars will of course, have already been detected by
conventional pulsar surveys. In order to find out the number of pulsars 
that {\bf only} the variance imaging can find, we need to take 
these previous major pulsar surveys into account.
We simulate pulsar surveys using the {\sc dosurvey} module of PSRPOPPY.
Pulsar surveys that cover sky areas of
interest include the Parkes Southern Pulsar Survey~\citep[PSPS;][]{mld+96}, the 
Parkes Multibeam Pulsar Survey~\citep[PMPS;][]{mlc+01}, the Swinburne Intermediate 
Latitude Survey~\citep[SILS;][]{ebv+01}, the High Time Resolution Universe~\citep[HTRU;][]{kjv+10}, 
the pulsar survey with the Arecibo L-band Feed Array~\citep[PALFA;][]{cfl+06}, 
the Arecibo all-sky 327\,MHz drift pulsar survey~\citep[AO327;][]{dsm+13} and
the Green Band telescope 350\,MHz drift-scan survey~\citep[GBT350;][]{blr+13}.
These simulated surveys are applied to the population of pulsars already 
`discovered' in variance images thus providing the number of pulsars
detected only by the variance imaging.

We carry out above simulations 100 times for normal pulsars and MSPs separately. 
This allows us to study DM, flux, duty cycle and sky distributions of new pulsars. 
Two future continuum surveys will be considered below, the EMU 
survey of ASKAP and an all-sky continuum survey of SKA-MID.

\subsection{The EMU survey on ASKAP}
\label{emu}

The EMU survey of ASKAP will make a deep ($\rm{rms}\sim10$\,$\mu$Jy/beam) radio 
continuum survey of the entire Southern sky at 1.3 GHz, covering the entire
sky south of $+30^{\circ}$ declination, with a resolution of 10\,arcsec~\citep{nha+11}. 
The EMU survey has a centre frequency of 1.3\,GHz, a total bandwidth of $B=300$\,MHz, and 
an integration time of $T=12$\,h for each pointing~\citep{nha+11}. The default time 
and frequency resolutions are 10\,s and 1\,MHz respectively, which will enable us 
to be sensitive to pulsars with shortest scintillation time-scales of 5\,s and smallest 
scintillation bandwidths of 0.5\,MHz. In order to investigate the ability of detecting 
new pulsars with different frequency resolutions, we also consider a frequency 
resolution of 0.2\,MHz, a factor of five higher than the default.

For the default EMU frequency and time resolutions of 1\,MHz and 10\,s, our simulations 
show that the variance imaging can detect 390 normal pulsars out of the 120,000 detectable 
normal pulsars and 167 MSPs out of the 41\,000 MSPs, of which 31 normal 
pulsars and 38 MSPs are new. 
If we increase the frequency resolution of EMU to 0.2\,MHz, the number of detection 
increases to 500 and 178 for normal pulsars and MSPs, respectively. However, 
increasing the frequency resolution does not increase the number of new pulsars 
that can be found.
This is because of the limited sensitivity of EMU and the fact that 
the sensitivity of variance imaging
decreases as we increase the frequency resolution and becomes lower than 
those of periodic searches.

\begin{figure*}
\center
\includegraphics[width=5in]{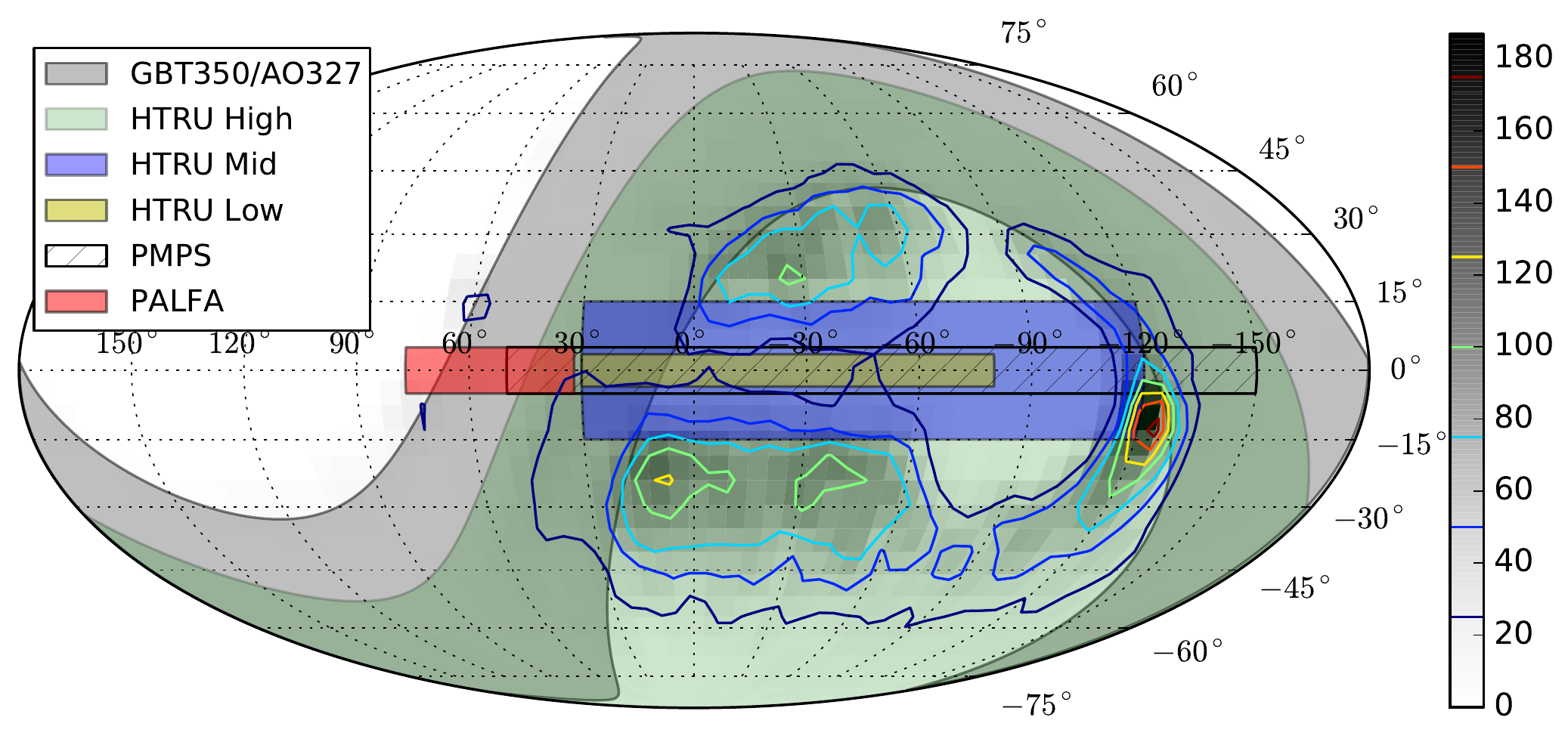}
\caption{Sky distribution of new pulsars that can be found using variance images with EMU.
The grey scale and contours show the number of new pulsars detected in 100 simulations.}
\label{sky_pro}
\end{figure*}

In Fig.~\ref{sky_pro}, we show the sky distribution of new pulsars (including both
normal pulsars and MSPs) detected in our 100 simulations. The sky is gridded into 
$5.5^{\circ}\times5.5^{\circ}$ size pointings, which roughly represents the beam size 
of ASKAP. The grey scale and contour show the number of pulsars detected in each 
pointing. 
We also show the sky coverage of several previous and on-going pulsar surveys 
with filled regions. 
Variance imaging of EMU will not be able to find new pulsars in regions 
that have been searched by deep pulsar surveys, e.g., regions searched by  
the PMPS, PALFA, GBT350 and AO327. 
At relatively high Galactic latitudes where 
only shallow surveys have been done EMU can find new pulsars. At Galactic 
longitude of $\sim120^{\circ}$, significantly more new pulsars can be detected 
by variance images. This is because of the much lower electron density towards 
that direction where we look into the space between two spiral arms, and also 
because this region has not been searched by deep surveys. On the other hand, 
at Galactic longitude of $\sim90^{\circ}$ where we look in the Carina-Sagittarius
arm no new pulsars are found because of the dense interstellar medium.

\begin{figure}
\center
\includegraphics[width=3.5in, trim=1.5cm 2.0cm 0.5cm 1.5cm]{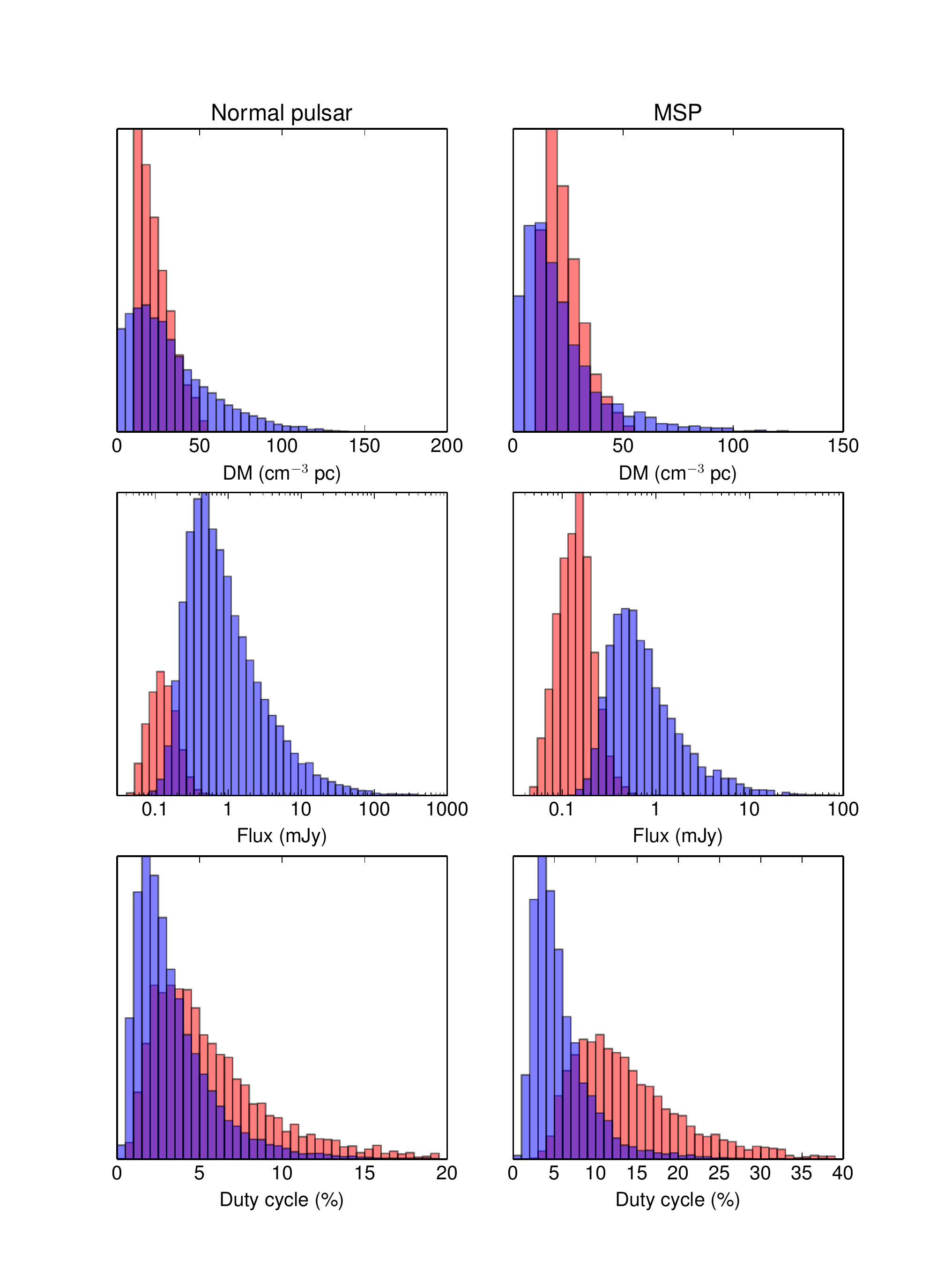}
\caption{Distributions of new normal pulsars and MSPs detected with the EMU survey (red), 
compared with pulsars discovered by periodic searches (blue). For normal pulsars, distributions 
of DM and duty cycle are normalised so that the area underneath is unity. Other distributions 
are not normalised in order to show the difference.}
\label{emu_hist}
\end{figure}

As shown in Fig.~\ref{sky_pro}, most new pulsars that can be detected by the variance imaging 
with EMU will be at relatively high Galactic latitudes. In Fig.~\ref{emu_hist}, 
we show distributions of DM, flux density and duty cycle of new normal pulsars 
and MSPs detected by EMU (red), and compare them with those of pulsars found by 
the HTRU high Galactic latitude survey (blue). Variance imaging of EMU will only 
be able to find new pulsars with DM between $\sim20$ and 50\,cm$^{-3}$\,pc because 
the scintillation of low DM pulsars will be too weak to detect and the scintillation of 
high DM pulsars will be too strong for EMU to resolve the flux variance. However, 
we can see that variance images of EMU will be more sensitive than the HTRU high
Galactic survey, and will be able to detect new faint pulsars with large duty
cycles.

\subsection{Continuum surveys with the SKA}
\label{ska}

A raft of different continuum surveys has been proposed for both SKA-MID
and SKA-LOW \citep[e.g.,][]{ps15}.
For surveys with SKA-LOW (from 50\,MHz up to 350\,MHz), even the
most nearby pulsars will be in the strong scintillation regime~\citep[e.g.,][]{bmj+16}. 
Variance imaging will then require very high
time and frequency resolutions with a subsequent reduction in sensitivity.
This implies that SKA-LOW continuum surveys will not yield many pulsars
via variance imaging and we do not consider them here. Rather, we focus 
on continuum surveys with SKA-MID in the 1.4\,GHz band. We note, however,
the SKA-LOW is a critical component for the conventional pulsar survey
expected to uncover many thousands for pulsars.

An all-sky continuum survey at 1.4\,GHz with a sensitivity of $\sim4$\,$\mu$Jy 
has been proposed for SKA-MID with an observing bandwidth of 810\,MHz, and 
an integration time is $\sim360$\,s \citep{ps15}. 
We assume a frequency resolution of 0.05\,MHz and a time resolution 
of 1\,s.
Following Section~\ref{var_sensitivity} we calculate the sensitivity of 
variance imaging constructed with such a continuum survey as a function of  
scintillation bandwidth and time-scale.
The observing bandwidth of 810\,MHz and frequency resolution of 0.05\,MHz 
allow us to detect pulsars with scintillation bandwidths between
$\sim0.025$\,MHz and $\sim400$\,MHz. The integration time of 300\,s and time 
resolution of 1\,s enable us to be sensitive to 
scintillation time-scale between 0.5 and 150\,s.
We also note that this imaging survey allows to use {\bf all} the SKA-MID
antennas and map the full field-of-view. This is not the case with the
conventional pulsar search which uses only the `core' antennas and has
a limited field-of-view~\citep[e.g.,][]{kbk+15}.

\begin{figure*}
\center
\includegraphics[width=5in]{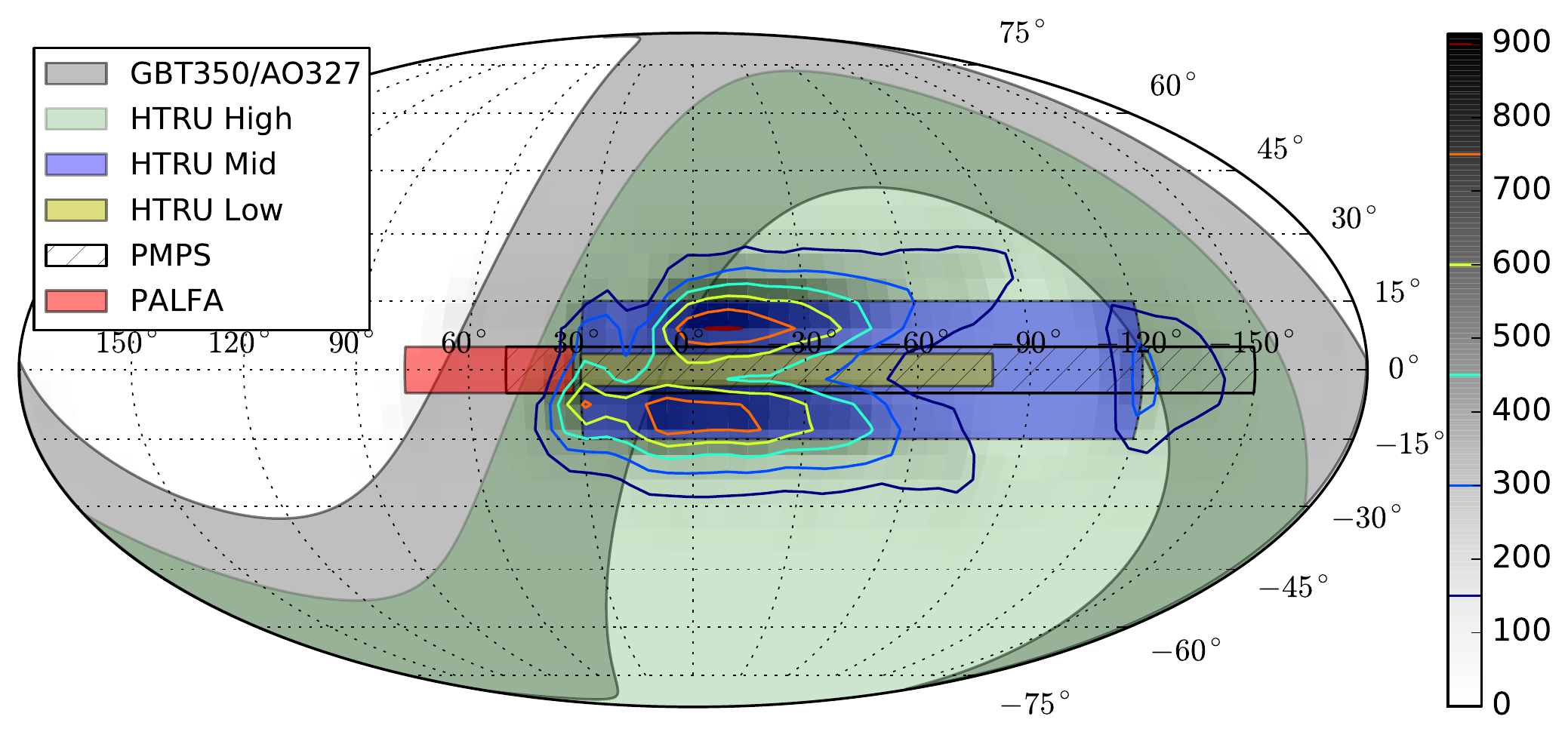}
\caption{Sky distribution of new pulsars that can be found using variance imaging with SKA-MID. 
The grey scale and contours show the number of new pulsars detected in 100 simulations.}
\label{sky_pro_ska}
\end{figure*}

\begin{figure}
\center
\includegraphics[width=3.5in, trim=1.5cm 2.0cm 0.5cm 1.5cm]{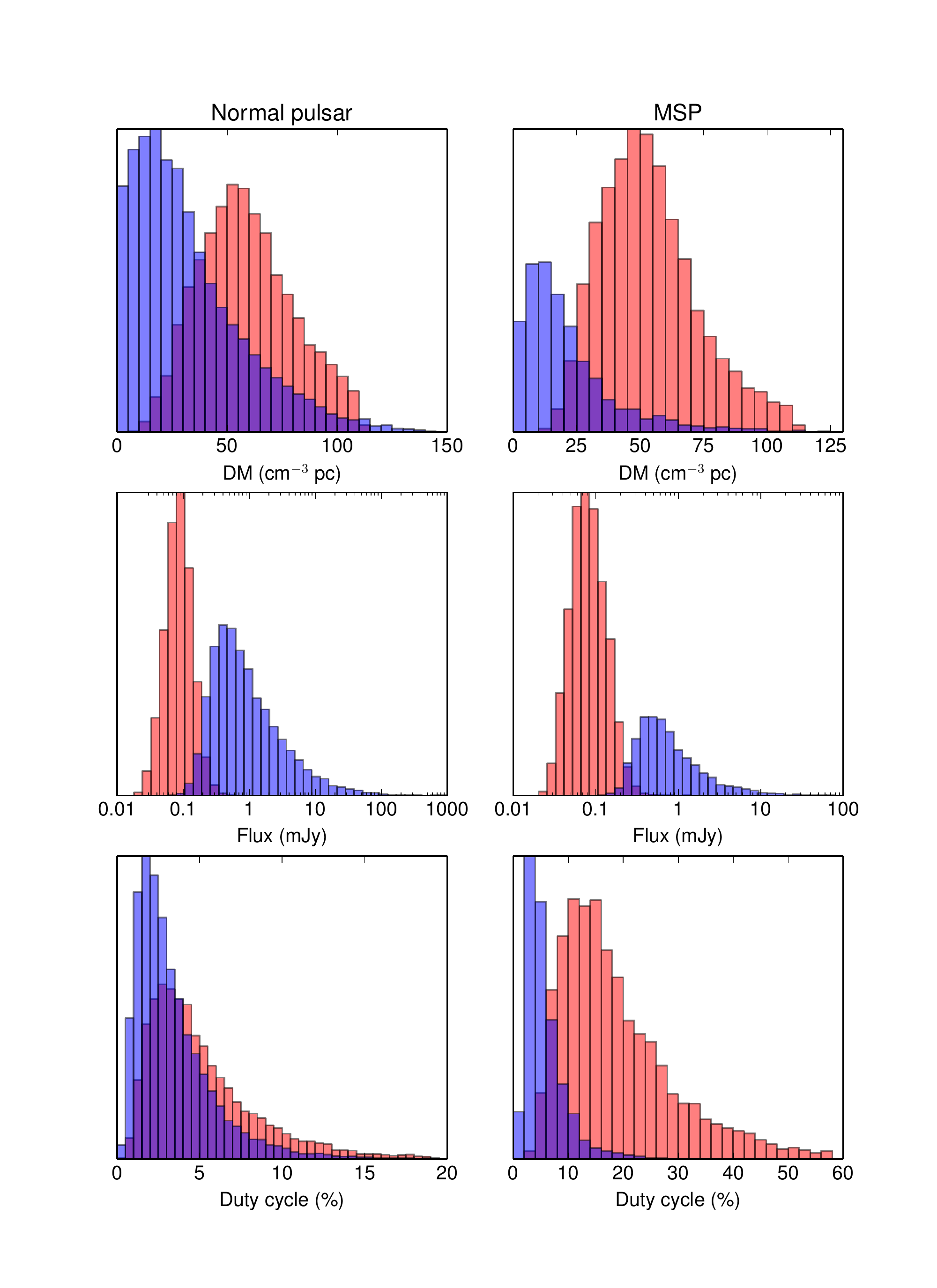}
\caption{Distributions of new normal pulsars and MSPs detected with the SKA-MID survey (red), 
compared with pulsars discovered by periodic searches (blue). Distributions are not normalized.}
\label{ska_hist}
\end{figure}

We separately simulate 100 populations of normal pulsars and MSPs and predict the number of new 
pulsars that can be detected in variance images. We find that an SKA-MID all-sky continuum 
survey can detect 671 normal pulsars and 207 MSPs in variance images of which 143 normal pulsars and 
113 MSPs are new. In Fig.~\ref{sky_pro_ska}, we show the sky distribution of new pulsars (including 
both normal pulsars and MSPs). In Fig.~\ref{ska_hist}, we show distributions of DM, flux density 
and duty cycle of new normal pulsars and MSPs detected by SKA-MID (red), and compare them with those 
of pulsars found by the HTRU high Galactic latitude survey (blue). Compared with Fig.~\ref{sky_pro} 
and \ref{emu_hist}, we can see that variance imaging with SKA-mid will be able to detect new pulsars 
with higher DM (up to $\sim110$\,cm$^{-3}$\,pc) because of the higher sensitivity and frequency 
resolution. On the other hand, variance imaging of SKA-MID will not be 
good at detecting low DM pulsars mostly because of the very short integration time. Again, 
variance imaging with SKA-MID can find pulsars with large duty cycles, especially for MSPs.
The sky distribution of new pulsar is different from that of EMU as we can detect pulsars with
higher DM and more pulsars can be found closer to the Galactic plane.

\section{Discussion}
\label{discussion}

\subsection{Limitations of the simulation}
Critical to our simulation is the computation of scintillation bandwidth
as a function of distance and galactic coordinates.
The electron-density model from \citet{ymw17} was used to 
predict the DM and we used a relation obtained by \citet{kmn+15} for the
DM dependence of observed $\tau_{\rm{sc}}$ values.
 
Different electron-density models and different ways of estimating 
$\tau_{\rm{sc}}$ can result in significantly different scintillation bandwidth 
and time-scales for the diffractive scintillation.
The widely used NE2001 electron-density model~\citep{cl02} predicts 
smaller $\tau_{\rm{sc}}$ and weaker scintillation (wider scintillation 
bandwidth) compared with \citet{ymw17}, 
and therefore allows the variance imaging technique to detect higher DM pulsars. Using the NE2001 model 
and leaving all other parameters fixed we find that EMU can detect $\sim90$ 
new normal pulsars and $\sim60$ new MSPs; the all-sky continuum survey of 
SKA-MID can detect $\sim230$ new normal pulsars and $\sim150$ new MSPs.

The population of potentially observable pulsars in the Galaxy is essential for 
predictions of detections for future continuum surveys.  
Recently, \citet{jk17} show that the decay of the inclination angle ($\dot{\alpha}$) between 
the magnetic and rotation axes plays a critical role in pulsar evolutions and predict 
a total Galactic population of 20\,000 normal pulsars beaming towards Earth. 
In our simulations, we used a Galactic population of 120,000 observable normal pulsars 
predicted by \citet{fk06}, which is much larger than the prediction of \citet{jk17}. 
However, both \citet{fk06} and \citet{jk17} produce similar results at the flux densities
under consideration here with their main differences arising at the
low flux densities available only to the conventional pulsar searches with
the SKA. \citet{jk17}, do however predict more pulsars with large duty cycles
and, potentially, variance imaging could verify this prediction.

Most MSPs are in binary systems, and some of these systems can be very compact which
allow us to test gravity theories in strong gravitational field. For MSPs in relativistic 
binary systems, the spin period is strongly modulated by the orbital period and these 
systems will be very difficult to detect through periodic searches. In our simulations, 
we did not include orbital period of pulsars and assumed all MSPs to be isolated. 
However, we clearly show that variance images detect more large duty-cycle MSPs, and 
therefore should be more sensitive to MSPs in relativistic binary systems than periodic searches. 

\subsection{Implications}
Our simulations show that variance imaging of the EMU survey of ASKAP should
find $\sim30$ new normal pulsars and $\sim40$ new MSPs, and an all-sky continuum survey
with SKA-MID should detect $\sim140$ new normal pulsars and $\sim110$ new MSPs.
Our results indicate that
\begin{itemize}
\item at 1.4\,GHz, variance imaging with future surveys will 
only be able to detect pulsars with DM below $\sim120$\,cm$^{-3}$\,pc and
hence are most likely to find pulsars at relatively high Galactic latitudes.
To detect high DM pulsars requires higher frequency surveys with
high sensitivity.
\item variance imaging with EMU has only limited sensitivity and
is unlikely to find a large number of new pulsars. However, it
is more sensitive than current pulsar surveys at high Galactic latitudes
in the southern hemisphere and so 
variance imaging with EMU provides us an efficient way to search for
pulsars at these high latitudes.
\item when compared with conventional pulsar searches, variance imaging is
better at identifying pulsars with large duty cycles, and therefore is 
a powerful tool for finding MSPs, whose duty cycles are normally larger than
normal pulsars. 
The spin-periods of MSPs in relativistic binary systems will be 
heavily modulated by their orbital velocity, which results in large duty 
cycles if the orbital motion is not corrected. Variance imaging can help us
identify such systems.
\item variance imaging with SKA-MID continuum surveys provides complementarity to the
conventional pulsar searches planned for SKA-MID and SKA-LOW.
\citet{kbk+15} predict that an all-sky pulsar survey with SKA-LOW can detect
$\sim900$ MSPs, and the DM distribution of these MSPs peaks at
$\sim160$\,cm$^{-3}$\,pc. Our simulations show that variance imaging
with SKA-MID all-sky continuum survey can detect $\sim210$ MSPs with DM 
below $\sim120$\,cm$^{-3}$\,pc. 
\item variance imaging can help us distinguish pulsars from other compact radio 
sources, but is less sensitive than continuum imaging and is
limited to relatively low DM pulsars. Variance imaging is only one
possible avenue for finding pulsars in continuum images; searches in
polarisation, rotation measure and circular polarisation are other options.
\end{itemize}

\subsection{Implementation}

The next stage for this project is to implement these ideas on
existing data and to plan for large-scale reduction of the future ASKAP
surveys. Much work is still needed to deal with the practicalities
of real data. Radio frequency interference needs to be mitigated.
It remains unclear whether or not deconvolution of the images will be
necessary and this will likely depend on snapshot u-v coverage.
A technique for filtering out sources with smoothly varying spectra (e.g., power-law) 
needs to be developed. In general, however, we expect the generation of variance 
images to be computationally cheap compared to the cost of generating 
the continuum images.

We also need to consider the false alarm threshold. In this paper this
is set, somewhat arbitrarily, to five per cent. Raising this level will result in the detection of more pulsars but will also increase the number
of potential candidates needing follow-up observations. In order to avoid this number becoming too large, other criteria can be used. For
example, if a radio source is identified with an optical galaxy or if
it is extended it can be ruled out as being a pulsar even if it shows
up in the variance image. The false alarm threshold will therefore be
survey specific and will depend on a raft of different inputs.

Finally, we note that continuum images with existing telescopes 
can already be used to detect pulsars~\citep[e.g.,][]{kca+98,kcc+00,ht99,ckb00,k00}. 
\citet{bmj+16} studied the variability of pulsars in the southern sky 
using 154 MHz Murchison Widefield Array imaging observations.
While the limited sensitivity of current all-sky continuum surveys makes
it difficult to discover new pulsars, ultra-deep images have been used to 
search for pulsars in the Galactic center~\citep[e.g.,][]{lc08,bdf+17}.
In order to detect pulsars in variance images, we need to have enough 
bandwidth, frequency and time resolution to resolve diffractive 
scintillation, and have good snapshot sensitivity and u-v coverage. 
This can already be achieved by recent continuum surveys~\citep{djb+16}, 
and can be combined with conventional pulsar searches to identify extreme 
pulsars.

\section{Conclusions}
\label{conclusion}

The optimum method for finding pulsars is through time-series analysis
which involves de-dispersion, Fourier transforms and harmonic summing.
These techniques have reaped 2500 pulsars to date, with the SKA expected
to find in excess of 10000 pulsars. Interferometric pulsar searches 
come with a heavy computational cost, and it is currently envisaged 
that the SKA survey can only make use of limited number of antennas 
and a limited field of view.

Other, complementary techniques for uncovering pulsars in
continuum surveys, which essentially come `for free' along
with the imaging, are therefore worth exploring. Such techniques include 
examining steep spectrum objects, searching in Rotation Measure space, 
searching in circular polarisation and through the scintillation properties 
of pulsars. We have examined the viability of finding pulsars via 
variance imaging and show that while the yield of pulsars from the EMU 
survey on ASKAP is small, surveys with SKA-MID can significantly add 
to the haul of MSPs in particular.

Although variance imaging with EMU and SKA-MID is only sensitive to pulsars 
with relatively low DM (less than $\sim120$\,cm$^{-3}$\,pc), they are equally 
sensitive to pulsars with different duty cycles, and are therefore better at 
identifying pulsars with large duty cycles than conventional pulsar searches.
Variance imaging is likely to be a powerful technique for finding pulsars in 
relativistic binary systems. 



\bibliography{ms}

\bsp	
\label{lastpage}
\end{document}